\def\preprint{1}      

\ifdefined\preprint
  \documentclass[preprint,review,12pt]{elsarticle}
\fi
\ifdefined\wordcount
  \documentclass[final,3p,times,twocolumn]{elsarticle}
\fi
\ifdefined\final
  \documentclass[final,3p,times,twocolumn]{elsarticle}
\fi

\usepackage{amsmath}
\usepackage{amssymb}
\usepackage{caption}
\usepackage{graphicx}
\usepackage{color}
\usepackage{subfloat}
\usepackage{subfigure}
\usepackage{booktabs}
\usepackage{caption}
\usepackage{xcolor}
\usepackage{tikz}
\usepackage{epstopdf}
\usepackage{color}
\usepackage{bm}

\usepackage{comment}
\usepackage[figurename=Fig.]{caption}
\usepackage{tikz}
\usetikzlibrary{positioning,spy}
\usepackage{stackengine}

\biboptions{sort&compress}

\begin{document}

\begin{frontmatter}

\title{Influence of preferential diffusion on the distribution of species  in lean ${\mathrm{H_2}}$-air laminar premixed flames at different equivalence ratios}

\author[NCL]{Frederick W Young}

\author[NCL]{Umair Ahmed\corref{cor1}}
\ead{umair.ahmed@newcastle.ac.uk}

\author[NCL]{Nilanjan Chakraborty}

\address[NCL]{School of Engineering, Newcastle University, Newcastle upon Tyne NE1 7RU,\\ United Kingdom}
\cortext[cor1]{Corresponding author:}

\begin{abstract}
The influence of equivalence ratio on preferential diffusion effects and the resulting changes in the distributions of major species and their reaction rates have been analysed based on 2D simulations of lean ${\mathrm{H_2}}$-air laminar premixed flames, at $\phi=0.4$ and $0.7$. The enhancements of burning rate, flame surface area, and stretch factor increase with decreasing equivalence ratio with the increase in stretch factor particularly prominent when the burning rate and flame area are evaluated based on normalised mass fraction variation of ${\mathrm{H_2}}$. The preferential diffusion effects have been demonstrated to lead to significant deviations of mass fractions of major species and their reaction rates from the corresponding 1D unstretched laminar premixed flame solution and local variations of equivalence ratio. This tendency is particularly strong for ${\mathrm{H_2}}$ among all the major species. Moreover, mass fractions of ${\mathrm{O_2}}$ and ${\mathrm{H_2O}}$ are found to assume super-adiabatic values at the super-adiabatic temperature zones and this trend is particularly strong for the $\phi=0.4$ case. It has been demonstrated that the deviations of major species mass fractions and their reaction rates from their corresponding 1D unstretched laminar premixed flame values arise principally due to preferential diffusion effects induced by relative focussing/defocussing of species and heat at the positively and negatively curved regions with the nature of the deviations being opposite to each other depending on the sign of the curvature. The variations of normalised species mass fractions of $\mathrm{O_2}$ and $\mathrm{H_2O}$ are found to be significantly affected by the local equivalence ratio within the 2D laminar flame with $\phi=0.4$ but these effects weaken with an increase in global equivalence ratio.
 
\end{abstract}


\end{frontmatter}
\ifdefined \wordcount
\clearpage
\fi

\section{Introduction}
Gas turbine engine technologies relying on lean premixed combustion of hydrogen-based fuels are promising in the pursuit of improved sustainability in power generation and propulsion sectors. No carbon dioxide is produced from the combustion of pure hydrogen, while limiting to lean conditions reduces the production of NOx and reduces the flame speed, which is important due to the propensity for flashback in premixed hydrogen flames. 
It is well-known that preferential diffusion rates of heat and species due to non-unity Lewis number can lead to thermo-diffusive instability in premixed flames under certain conditions \cite{Sivashinsky1977, Matalon1982}. In the case of hydrogen-air mixtures, lean flames can be characterised by an effective Lewis number, $Le_{eff}$, much smaller than unity. Effective Lewis number is essentially a weighted average of the Lewis numbers of the reactants, with greater weighting given to the deficient reactant \cite{Matalon2003}. Thermodiffusive instabilities arise in the cases with sufficiently small values of the effective Lewis number \cite{Sivashinsky1977, Matalon1982, Matalon2003, Bechtold2001}. According to Bechtold and Matalon \cite{Bechtold2001}, the effective Lewis number $Le_{eff}$ of ${H_2}$-air is defined as: $Le_{eff}=1+[(Le_{\mathrm{O_2}}-1)+(Le_{\mathrm{H_2}}-1)\Xi]/(1+\Xi)$ where $\Xi=1+Ze(1/\phi-1)$, with $Le_{\mathrm{O_2}}>1.0$ and $Le_{\mathrm{H_2}}<<1.0$; the Lewis numbers of ${\mathrm{H_2}}$, and ${\mathrm{O_2}}$ and $Ze$ being the Zel'dovich number. This implies that $Le_{eff}$ is expected to decrease with decreasing equivalence ratio ($\phi$). Therefore, the likelihood of thermo-diffusive instability increases as $\phi$ decreases for lean ${\mathrm{H_2}}$-air flames. \par
In previous studies of laminar two-dimensional lean premixed hydrogen-air flames \cite{Altantzis2012, Berger2019, Creta2020, Howarth2022, Berger2022, Berger2022b, Berger2022a}, thermo-diffusive instabilities have been shown to amplify perturbations applied to the flame front, leading to the development of a highly wrinkled flame, with high flame surface area and burning rate. For developed wrinkled flames a broad range of curvature values are observed along the flame front, and for sufficiently lean flames ($\phi << 1$) a comparison of heat release statistics between positive and negatively curved regions reveals local extinction in flame cusps characterised by a strongly negative curvature (here, flame normal is defined to point into the unburned gas and flame surface elements convex (concave) towards the reactants have positive (negative) values of curvature).
The empirical evidence suggests a strong influence of curvature (and by extension strain and stretch rates) on the local flame behaviour, as suggested by Markstein \cite{Markstein1964}. 
\par
Recently, it has been shown based on three-dimensional Direct Numerical Simulations (DNS) of turbulent hydrogen-air flames that the choice of the species used to define the flame surface can significantly influence the evaluation of the flame surface area \cite{Klein2020}. In the context of Reynolds Averaged Navier-Stokes (RANS) and Large Eddy Simulations (LES) turbulent combustion models often make use of a reaction progress variable to quantify the conversion of fresh reactants to burned products. Commonly used flamelet-based closures \cite{ThierryPoinsot2005,deSwart2010,Ihme2012} need the knowledge of a reaction progress variable in terms of normalised mass fractions of major species so that the reaction progress variable increases monotonically from $0.0$ in the unburned gases to $1.0$ in fully burned products. The choice of the reaction progress variable depends on the distributions of major species and their responses to thermo-diffusive instability in lean ${\mathrm{H_2}}$-air flames which is yet to be fully understood. Therefore, it is necessary to analyse the behaviour of normalised major species mass fractions in lean ${\mathrm{H_2}}$-air flames for different equivalence ratios to understand the influence of thermo-diffusive instability on species distributions within the flame front. In this respect, the main objectives for this work are: (i) to investigate the variation of major species across lean hydrogen-air laminar premixed flames to ascertain the choice of the species relevant for defining an appropriate progress variable under fuel-lean conditions, (ii) to investigate the influence of preferential diffusion on the species distribution and (iii) to investigate the influence of the choice of the species on the evaluations of flame surface area and the stretch factor, $I_0$ \cite{Bray1990}. To achieve the aforementioned objectives, numerical simulations for two-dimensional laminar premixed hydrogen-air flames at two equivalence ratios, $\phi=0.4$ and $0.7$, have been performed with a simulation setup similar to that of Berger et al. \cite{Berger2019}. It was demonstrated by Berger et al. \cite{Berger2022} using dispersion relations that ${\mathrm{H_2}}$-air flames with $\phi=0.4$ are thermo-diffusively unstable (with $Le_{eff}\approx 0.34$ \cite{Berger2022}), whereas ${\mathrm{H_2}}$-air flames with $\phi=0.7$ (with $Le_{eff}\approx 0.5$ \cite{Berger2022}) are nominally thermo-diffusively neutral to stretch effects \cite{Chen2000}, and preferential diffusion effects are expected to be weaker in the $\phi=0.7$ case than in the $\phi=0.4$ case \cite{Berger2022}. Thus, these two equivalence ratios have been chosen to compare and contrast the effects of $\phi$ variation in lean ${\mathrm{H_2}}$-air premixed combustion.
\par
The rest of the paper is organised as follows. In the next section details of the numerical simulations are provided, which is followed by the results and the corresponding discussion. Finally, the findings are summarised and the conclusions are drawn in the final section of the paper.

\section{Numerical simulations\label{sec:numsims}} \addvspace{10pt}
A fully compressible direct numerical simulation code, SENGA2 \cite{Cant2013} is used to perform the simulations. The code solves conservation equations for mass, momentum, energy and species in time and space. A 10th-order central finite difference scheme (gradually reducing to fourth-order at non-periodic boundaries) is used to calculate spatial derivatives. Time advancement is performed using a low-storage fourth-order explicit Runge-Kutta scheme. In the simulations, the thermophysical properties such as specific heat, are assumed to be temperature-dependent and are evaluated using CHEMKIN polynomials. Molecular transport is implemented using a mixture-averaged approach, which includes the Soret and Dufour effects, and is used for the calculation of viscosity, species diffusivity, thermal conductivity and thermal diffusion ratio. The skeletal chemical mechanism for hydrogen-air combustion proposed by Burke et al. \cite{Burke2012}, which has 9 species and 23 reversible reactions, has been employed for the simulations considered here. \par 
Simulations of 2D flames at two different equivalence ratios have been performed, $\phi=0.4$ and $\phi=0.7$ under atmospheric conditions with an unburned gas temperature of $T_u=300$K. 
A domain length of $L_x = 450\delta_{th}$ and a domain width of $L_y = 260 \delta_{th}$ is used in both cases. A large domain width is chosen to allow for the development of a sufficient number of flame structures within the domain to ensure that domain size-independent behaviour is observed, whilst a large domain length is chosen to allow sufficient room for the flame to propagate along the domain, considering the expected enhancement of flame propagation speed due to the emergence of flame instabilities \cite{Berger2022a}. The domain for $\phi=0.4$ has been discretised using 6720 $\times$ 3840 points, whilst for $\phi=0.7$ the domain has been discretised using 6640 $\times$ 3840 points. The resolution requirement for both cases is determined by ensuring at least 14 grid points are used to resolve the thermal flame thickness, $\delta_{th}$. 
Periodic boundary conditions have been applied in the $y$-direction, with an inflow boundary condition imposed at $x=0$ and a non-reflecting outflow boundary condition imposed at $x=L_x$. Navier-Stokes Characteristic Boundary Conditions proposed by Lodato et al. \cite{Lodato2008} are used for specifying the conditions at the inflow and outflow boundaries within the simulations.
In order to trigger the unstable flame response for the analysis of the developed flame, an initial sinusoidal perturbation is applied in the $y$-direction to the initial fields, which themselves are taken from the steady-state 1D unstretched laminar flames. Low amplitude random numerical noise is imposed on the wave amplitude to ensure the development of non-repeating, asymmetric flame structures. The simulations are run until the burning rate and flame area has reached a quasi-steady state.

\begin{figure*}[h!] 
  \centering
  \includegraphics[width=410pt]{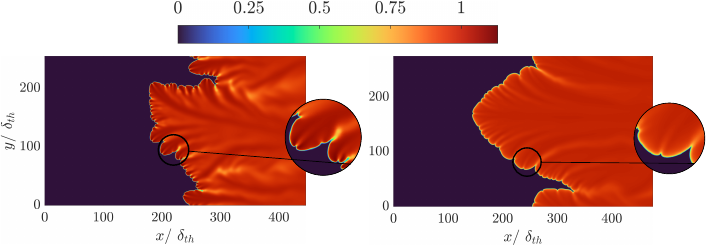}
  \vskip0.5em
  \caption{\footnotesize Distributions of $c_T$ fields for the developed flames with (left) $\phi$ = 0.4 and (right) $\phi$ = 0.7}
  \label{temp_fields}
\end{figure*}

\section{Results \& Discussion\label{sec:extext}} \addvspace{10pt}
Figure \ref{temp_fields} shows the distributions of instantaneous non-dimensional temperature $c_T=(T-T_u)/(T_{ad}(\phi)-T_u)$ fields for the two cases considered. Figure \ref{temp_fields} shows a strongly corrugated flame front for both cases. This flame wrinkling gives rise to a significant increase in the flame area. In the case of $\phi=0.4$ case, super-adiabatic flame temperatures (i.e., $c_T>1.0$) can be seen where the flame is convexly curved towards the unburned mixture, whereas super-adiabatic temperatures are rarely obtained in the $\phi=0.7$ case. In both cases, comparatively small temperature values (especially $c_T<1.0$ in the burned gas region) are obtained in the wake of the concavely curved regions. For flames with $Le_{eff}<1$ focusing of reactants takes place faster than the defocusing of heat at the positively curved zones \cite{Chakraborty2005,Chakraborty2009}. This gives rise to the simultaneous occurrence of high reactant concentration and temperature, leading to high fuel consumption and heat release rates at the positively curved zones in the flames with  $Le_{eff}<1$. This mechanism strengthens with decreasing $Le_{eff}$ \cite{Chakraborty2005,Chakraborty2009}. Thus, super-adiabatic temperatures are more prevalent in the $\phi=0.4$ case than in the $\phi=0.7$ case. The negatively curved regions in the $Le_{eff}<1$ flames are associated with high defocusing of reactants and weak focusing of heat, and this combination reduces the fuel consumption and heat release rates in these zones \cite{Chakraborty2005,Chakraborty2009}, which reduces the temperature at the negatively curved zones in both cases but this trend strengthens with decreasing $Le_{eff}$ \cite{Chakraborty2005,Chakraborty2009}. Hence, low temperature wakes in the burned gas are more prominent in the $\phi=0.4$ case than in the $\phi=0.7$ case. Figure \ref{temp_fields} also shows that positively curved flame segments are interrupted by sharp cusps with strong negative curvatures for flames with both values of $\phi$. All of the features of flame wrinkling mentioned above are consistent with previous findings for $Le_{eff}<1$ flames \cite{Chakraborty2009,Rasool2021,Berger2019,Berger2022a,Howarth2022}.

The thermo-diffusive instability enhances the burning rate \cite{Sivashinsky1977, Matalon1982, Berger2019,Berger2022a,Howarth2022}, which can be quantified in terms of the burning velocity $S_f$, defined in two dimensions in the following manner based on species $\alpha$:
\begin{equation}
  S_f = \frac{1}{\rho_u (Y_{\alpha,b}-Y_{\alpha,u}) L_y} \int_{0}^{Ly} \int_{0}^{Lx} \dot{\omega}_{\alpha} \ dx \ dy,
  \label{S_f}
\end{equation}
where $\rho_u$ is unburned gas density, $L_y$ is the span-wise domain dimension and $\dot{\omega}_{\alpha}$ is the reacion rate of species $\alpha$. It has been found that $S_f/S_L$ assumes greater values ($S_f/S_L \approx 3.5$ and $2.0$ in the $\phi=0.4$ and $0.7$ cases, respectively) in the $\phi=0.4$ case than in the $\phi=0.7$ case as a result of stronger thermo-diffusive effects induced by preferential diffusion in the $\phi=0.4$ case, which results in a larger increase in reactivity relative to that of the 1D unstretched laminar flame. This is consistent with previous findings by Berger et al. \cite{Berger2022a,Berger2022}. \par

\begin{figure}[h!] 
  \centering
  \includegraphics[width=\columnwidth]{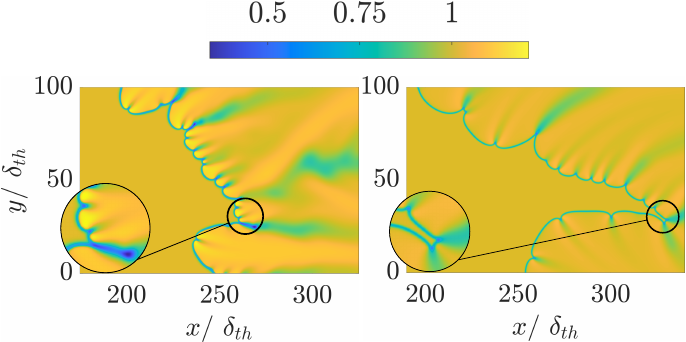}
  \caption{\footnotesize Magnified views of local normalised equivalence ratio $\phi$ in the 2D $\phi=0.4$ (left) and $\phi=0.7$ (right) laminar flame cases.}
  \label{isolines}
\end{figure}

\begin{figure*}[!ht]
  \centering        
   \includegraphics[width=\columnwidth]{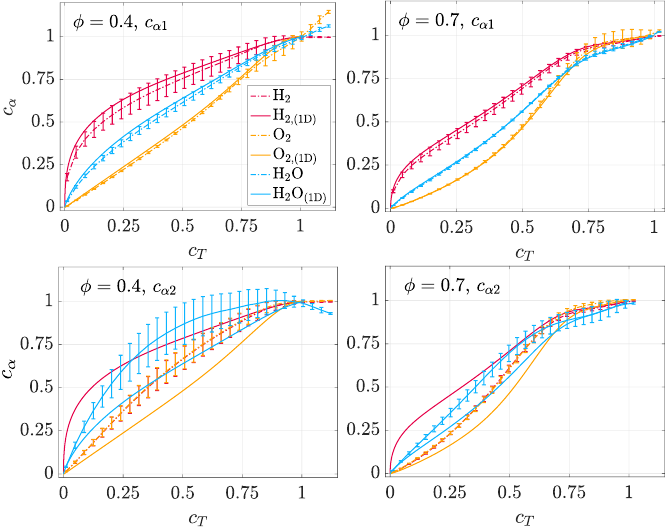}
  \vskip0.5em
  \caption{\footnotesize Mean values of $c_{\alpha 1}$ and $c_{\alpha 2}$ based on $\mathrm{H_2}$, $\mathrm{O_2}$ and $\mathrm{H_2O}$ mass fractions conditioned upon $c_T$ for both the original definition of $c_{\alpha}$ (denoted as $c_{\alpha 1}$), and the local mixture fraction dependent definition of $c_{\alpha}$ (denoted as $c_{\alpha 2}$). The error bars indicate standard deviations of each $c_{\alpha}$ conditioned upon $c_T$.}
  \label{c_avg}
\end{figure*}

In addition to the enhancement of burning velocity and the curvature dependence of temperature, the preferential diffusion effects alter the distribution of the species in the lean $\mathrm{H_2}$-air flames. This variation of species can lead to variation in local equivalence ratio, $\phi=(\xi(1-\xi_{st}))/((1-\xi)\xi_{st})$ (where $\xi$ is the local mixture fraction and $\xi_{st}$ is the stoichiometric mixture fraction), as reported by Berger et al. \cite{Berger2022a}. Here, the mixture fraction is defined as $\xi=(\beta-\beta_O)/(\beta_f-\beta_O)$, where $\beta = 2Y_C/W_C + 0.5Y_H/W_H-Y_O/W_O$, $\beta_f=(2a+0.5b-c)/W_{C_a H_b O_c}$ (where $a=0$, $b=2$ and $c=0$), and $\beta_O=-Y_{O \infty}/W_O$ with $Y_{O \infty}$ being the elemental mass fraction of oxygen in the pure oxidiser stream, and $Y_{\alpha}$ and $W_{\alpha}$ are mass fraction and molecular mass of the element or species $\alpha$, respectively \cite{Bilger1980}.
Figure \ref{isolines} shows the local variation of $\phi$ across the flame for the two cases considered here, where the local $\phi$ values have been normalised by the global equivalence ratio for each case ($\phi = 0.4$ and $0.7$ respectively) to aid comparison. It can be seen from Fig. \ref{isolines}  that the local variation in $\phi$ across the flame is much higher in the case of $\phi=0.4$ than in the $\phi=0.7$ flame which is consistent with the earlier findings of Berger et al. \cite{Berger2022a}. The observed differences between the positively and negatively curved regions in Fig. \ref{isolines} for $\phi=0.4$ case are indicative of the strong preferential diffusion effects induced by flame curvature, which affect the local fuel consumption rate and cause significant variations in local equivalence ratio. While a departure from the global equivalence ratio (and corresponding mixture fraction value) is expected across the flame due to the conversion of reactants to products via intermediate species, significant differences in local equivalence ratio are observed between positive and negative curvature regions for both cases, with lower $\phi$ regions observed adjacent to negatively curved flame segments, and higher $\phi$ regions found adjacent to positive curvature flame segments. This is especially visible on the burned gas side of the flame for $\phi=0.4$, where local equivalence ratios are significantly greater than the global value of $\phi=0.4$ behind positively curved segments. At positive curvatures, a relative increase in equivalence ratio with higher reactivity supports the assertion that positively curved regions display enhanced burning rates. Conversely, low equivalence ratio values seen in negatively curved regions indicate that these regions have smaller concentrations of fuel. The combination of strong defocusing of reactants and weak focusing of heat leads to local flame extinction in the negatively curved cusps. This local extinction behaviour in the negatively curved cusps for $\phi=0.4$ was previously reported by Berger et al. \cite{Berger2022a}.

\begin{figure}[!ht]
  \centering        
  \includegraphics[width=0.55\columnwidth]{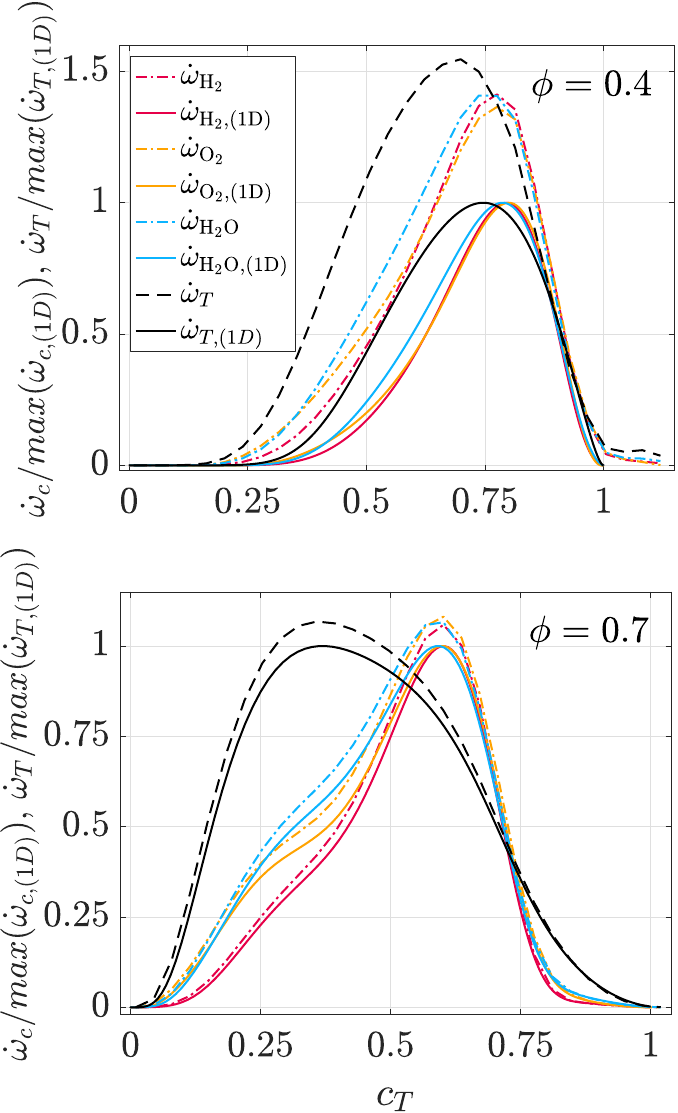}
  \vskip0.5em
  \caption{\footnotesize Mean values of normalised reaction rate of $c_{\alpha 1}$ (i.e., $\dot{\omega}_{c}$) based on $\mathrm{H_2}$, $\mathrm{O_2}$ and $\mathrm{H_2O}$ mass fractions, conditioned upon $c_T$. Normalisation of $\dot{\omega}_{c}$ and $\dot{\omega}_T$, has been performed using the maximum values corresponding to unstretched 1D laminar premixed flames.}
  \label{rrte_overall}
\end{figure}

 \par

In order to examine the preferential diffusion effects on the reaction zone structure, the variations of the mean values of the progress variable defined using two definitions, $c_{\alpha_1}=(Y_\alpha-Y_{\alpha,u})/(Y_{\alpha,b}-Y_{\alpha,u})$ and $c_{\alpha_2}=(Y_\alpha-Y_{\alpha}(\xi)_u)/(Y_{\alpha}(\xi)_b-Y_{\alpha}(\xi)_u)$, based on major species mass fractions and their normalised reaction rates (i.e.,  $\dot{\omega}_{c}=\dot{\omega}_{\alpha}/(Y_{\alpha,b}-Y_{\alpha,u})$ for $\alpha=\mathrm{H_2},\mathrm{O_2}$ and $\mathrm{H_2O}$) conditioned upon $c_T$ are shown in Fig. \ref{c_avg} and Fig. \ref{rrte_overall}, respectively for both $\phi=0.4$ and $0.7$. The mean values of the normalised heat release rate, $\dot{\omega}_T$, conditioned upon $c_T$ are also shown in Fig. \ref{rrte_overall}. The profiles corresponding to 1D unstretched laminar flames are also plotted in Fig. \ref{c_avg} and Fig. \ref{rrte_overall}, and error bars in Fig. \ref{c_avg} indicate the standard deviation of each averaged data point. Normalisation of $\dot{\omega}_{c}$ and $\dot{\omega}_T$, has been performed using the maximum values obtained for 1D unstretched laminar flames. For both Fig. \ref{c_avg} and Fig. \ref{rrte_overall}, data from across the entire domain has been considered in the averaging process. \par

It can be seen from Fig. \ref{c_avg} that $c_{\alpha 1}$ based on $\mathrm{H_2}$ mass fraction shows the highest standard deviations of the different species used to define $c_{\alpha 1}$, for both equivalence ratios. This is a consequence of the small Lewis number value of $\mathrm{H_2}$, which gives rise to a considerable scatter of $c_{\alpha 1}$ based on $\mathrm{H_2}$ mass fraction for a given value of $c_T$ due to preferential diffusion effects. Moreover, the mean profile of $c_{\alpha 1}$ based on $\mathrm{H_2}$ mass fraction significantly deviates from the corresponding 1D unstretched laminar flame profile for the $\phi=0.4$ case but this trend is much weaker in the $\phi=0.7$ case due to weakened preferential diffusion effects in the latter case. Figure \ref{c_avg} shows that the profiles for $c_{\alpha 1}$ based on $\mathrm{O_2}$ and $\mathrm{H_2O}$ mass fraction agree reasonably well with the corresponding 1D unstretched laminar flame profiles, apart from the behaviour in the super-adiabatic region ($c_T > $ 1) for the $\phi = 0.4$ flame. In the super-adiabatic region, $c_{\alpha 1}$ based on $\mathrm{O_2}$ and $\mathrm{H_2O}$ mass fractions assume values greater than unity, indicating the enhanced consumption (generation) of oxygen (water) in the 2D laminar $\phi=0.4$ premixed laminar flame in comparison to the corresponding 1D unstretched laminar flame. This can be substantiated by the non-zero values of $\dot{\omega}_{c}$ for $c_T > 1.0$ in the $\phi$ = 0.4 case. For the purpose of modelling, the reaction progress variable is usually bound between 0.0 and 1.0 and super-unity values of reaction progress variables based on  $\mathrm{O_2}$ and $\mathrm{H_2O}$ mass fractions indicate that a standard reaction progress variable definition based on $\mathrm{O_2}$ and $\mathrm{H_2O}$ mass fractions may not be suitable for thermo-diffusively unstable lean $\mathrm{H_2}$-air flames.\par
The standard deviations of all $c_{\alpha 1}$ profiles for the $\phi=0.7$ case are smaller than those for the $\phi=0.4$ case, which is a consequence of the reduction in the variability of the $c_{\alpha 1}$ values in comparison to the $\phi = 0.4$ case.
It can further be seen from Fig. \ref{rrte_overall} that qualitative behaviours of $\dot{\omega}_{c}$ and $\dot{\omega}_{T}$ for the 2D laminar flames are mostly in agreement with those of the corresponding 1D unstretched laminar flames. However, significant quantitative differences can be seen in the profiles of $\dot{\omega}_{c}$ and $\dot{\omega}_{T}$ between the 2D case and the corresponding 1D unstretched laminar flame case for $\phi=0.4$ in Fig. \ref{rrte_overall}, indicating an enhancement of heat release rate (e.g., there is an almost $50 \%$ increase in $\dot{\omega}_T$) and $\dot{\omega}_{c}$ values for each species mass fraction, highlighting the tendency of thermo-diffusive instabilities to enhance reactivity in lean $\mathrm{H_2}$-air flames. This is consistent with the previous findings by Berger et al. \cite{Berger2022a}. \par
The profiles shown in Fig. \ref{c_avg} corresponding to $c_{\alpha 2}$ are markedly different to those observed for $c_{\alpha 1}$ (as well as the profiles for the corresponding 1D unstretched laminar flames), which is a result of equivalence ratio variation within the flame, as shown in Fig. \ref{isolines}. Profiles for $c_{\alpha 2}$ based on H$_2$ and O$_2$ are very similar to each other at both equivalence ratios, whereas significant differences are observed between the $c_{\alpha 1}$ profiles for H$_2$ and O$_2$. Notably, $c_{\alpha 2}$ for O$_2$ is bounded between 0.0 and 1.0 as a result of mixture fraction dependence, which, as previously mentioned, is an important feature when selecting a suitable reaction progress variable. 
However, for $\phi=0.4$, relatively large standard deviations are seen in $c_{\alpha 2}$ profiles, particularly for $c_{\alpha 2}$ based on $\mathrm{H_2O}$, which also displays clear non-monotonic behaviour, reaching a value close to 1.0 at $c_T \approx 1$, before decreasing at $c_T>1$. This implies that the inclusion of mixture fraction dependence makes the variation of normalised $\mathrm{H_2O}$ mass fraction with $c_T$ non-monotonic and this may not be suitable for the definition of reaction progress variable. The standard deviations of $c_{\alpha 2}$ based on $\mathrm{O_2}$ mass fraction show a moderate increase when compared with the corresponding values for $c_{\alpha 1}$ whilst the standard deviations for $c_{\alpha 2}$ based on $\mathrm{H_2}$ mass fraction remain approximately the same as that obtained for $c_{\alpha 1}$. \par
\begin{figure*}[h!]
  \centering        
  \includegraphics[width=0.75\columnwidth]{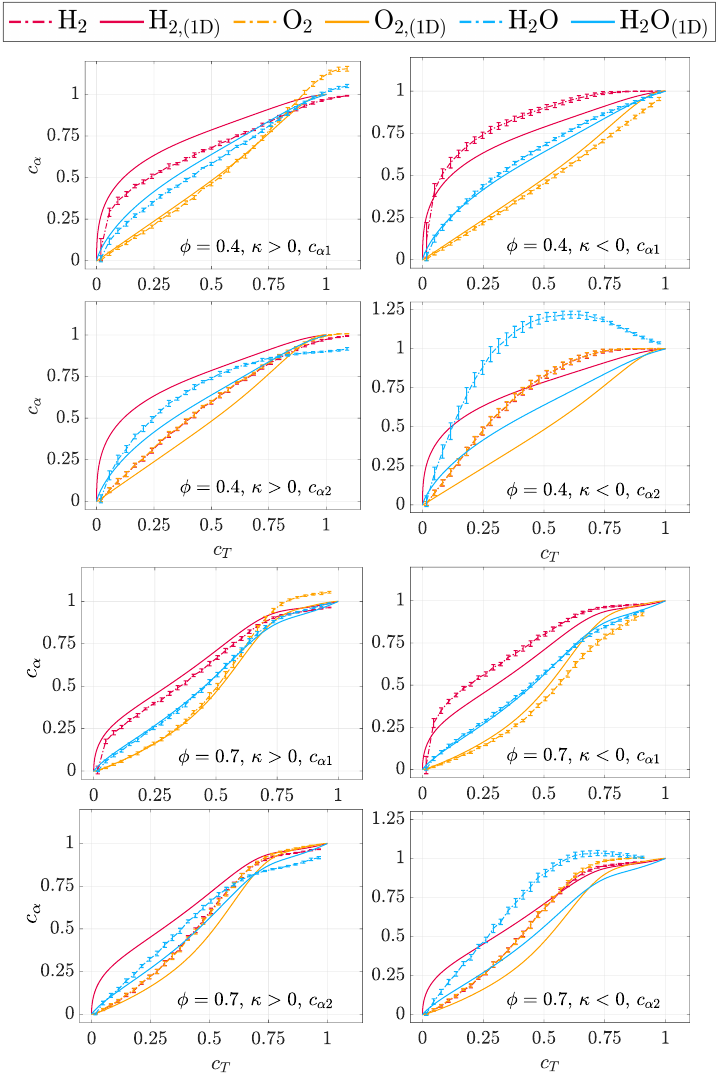}
  \vskip0.5em
  \caption{\footnotesize Mean values of $c_{\alpha}$ based on $\mathrm{H_2}$, $\mathrm{O_2}$ and $\mathrm{H_2O}$ mass fractions conditioned upon $c_T$ for positively (i.e., $\kappa>0$) and negatively (i.e., $\kappa<0$) curved regions for both the original $c_{\alpha}$ definition ($c_{\alpha 1}$) ($1^{st}$ row), and the local mixture fraction dependent $c_{\alpha}$ definition ($c_{\alpha 2}$) ($2^{nd}$ row). The error bars indicate standard deviations of $c_{\alpha}$ conditioned upon $c_T$.}
  \label{c_avg_profiles}
\end{figure*}
\begin{figure*}[h!]
  \centering        
  \includegraphics[width=0.9\columnwidth]{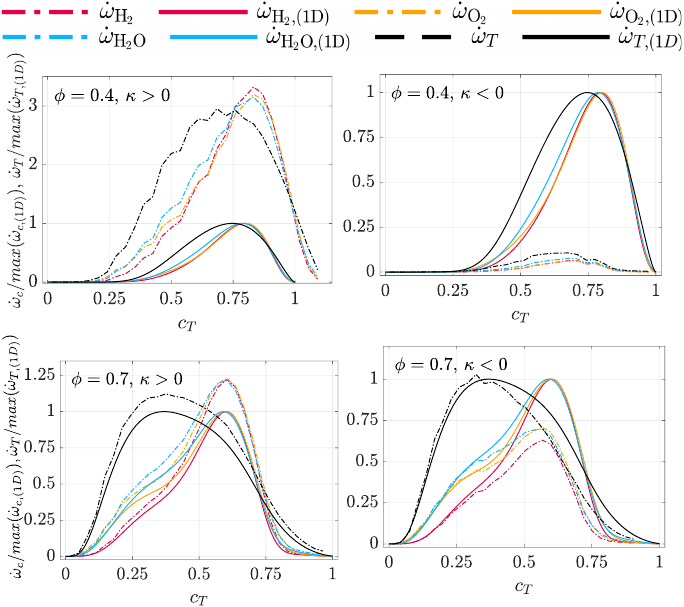}
  \vskip0.5em
  \caption{\footnotesize Mean values of normalised reaction rate of $c_{\alpha 1}$ (i.e., $\dot{\omega}_{c}$) based on $\mathrm{H_2}$, $\mathrm{O_2}$ and $\mathrm{H_2O}$ mass fractions conditioned upon $c_T$ for positively (i.e., $\kappa>0$) and negatively (i.e., $\kappa<0$) curved regions. Normalisation of $\dot{\omega}_{c}$ and $\dot{\omega}_T$, has been performed using the maximum values corresponding to unstretched 1D laminar premixed flames.}
  \label{rrte_profiles}
\end{figure*}
The departures of the 2D averaged profiles for $c_{\alpha 1}$ and $c_{\alpha 2}$ along with $\dot{\omega}_{c}$ from the corresponding 1D unstretched laminar flame profiles shown in Fig. \ref{c_avg} and Fig. \ref{rrte_overall} have been explored further by plotting mean values of $c_{\alpha 1}$ and $c_{\alpha 2}$ based on major species mass fractions and their normalised reaction rates conditioned upon $c_T$ for positive and negatively curved regions in Fig. \ref{c_avg_profiles} and Fig. \ref{rrte_profiles} respectively. A comparison between Fig. \ref{c_avg} and Fig. \ref{c_avg_profiles} reveals that conditioning on curvature leads to a significant reduction in the standard deviations of $c_{\alpha}$ for both $\phi=0.4$ and $0.7$ cases. This reduction in standard deviations when conditioning on curvature applies to both $c_{\alpha 1}$ and $c_{\alpha 2}$ profiles and implies that the preferential diffusion due to flame curvature is principally responsible for the variability of $c_{\alpha}$ values for a given value of $c_T$. Therefore, a much narrower distribution of $c_{\alpha 1}$ and $c_{\alpha 2}$ about their conditional mean values on $c_T$ is seen in Fig. \ref{c_avg_profiles} than in Fig. \ref{c_avg}. The profiles of mean values of $c_{\alpha 1}$ and $c_{\alpha 2}$ based on H$_2$ mass fraction conditional on curvature signs deviate significantly from the corresponding 1D unstretched laminar flame profiles, reflecting that curvature induces significant differences in the local flame structure due to preferential diffusion effects. However, the tendency for the 2D profiles to diverge from the corresponding 1D laminar profiles is weaker for $c_{\alpha 1}$ based on $\mathrm{O_2}$ and $\mathrm{H_2O}$ mass fractions in comparison to that of  $c_{\alpha 1}$ based on $\mathrm{H_2}$ mass fraction.
The profiles of mean values of $c_{\alpha 1}$ and $c_{\alpha 2}$ based on $\mathrm{H_2}$ mass fraction exhibit smaller magnitudes than the corresponding 1D unstretched laminar values in regions of positive curvatures for both equivalence ratios. The opposite is true for the regions with negative curvature, with mean $c_{\alpha 1}$ and $c_{\alpha 2}$ values based on ${\mathrm{H_2}}$ mass fraction being greater than the corresponding 1D unstretched laminar values across the vast majority of the flame. These trends of mean values of $c_{\alpha1}$ and $c_{\alpha2}$ based on $\mathrm{H_2}$ mass fraction are consistent with high (low) equivalence ratios at positively (negatively) curved zones, as shown in Fig. \ref{isolines}. \par

In Fig. \ref{rrte_profiles}, the profiles for $\dot{\omega}_{c}$ and $\dot{\omega}_T$ at $\phi$ = 0.4 support the assertion that local extinction occurs at negative curvature regions, with the reaction rate profiles for the negatively curved zones roughly an order of magnitude smaller than the corresponding 1D unstretched laminar flame values. Conversely, a significant enhancement of the reaction rates in positive regions is evident from Fig. \ref{c_avg_profiles}, which shows an approximately three-fold increase in the 1D unstretched laminar flame values in the $\phi=0.4$ case. In the $\phi=0.7$ case, the mean values of $\dot{\omega}_{c}$ conditional upon $c_T$ show much smaller variation from the corresponding 1D unstretched laminar flame values, where slightly smaller (greater) values are observed than the 1D unstretched laminar flames values for negative (positive) $\kappa$. \par
The profiles corresponding to $c_{\alpha 2}$ are significantly different from both those seen for $c_{\alpha 1}$ as well as the 1D unstretched laminar flame profiles conditioned upon signs of curvature. The non-monotonic behaviour of $c_{\alpha 2}$ based on H$_2$O is clearly explained by the behaviour in negative curvature regions where large overshoots of $c_{\alpha 2}$ are observed, with a maximum average value of $c_{\alpha 2}$ based on H$_2$O mass fraction close to $1.25$. Furthermore, as in Fig. \ref{c_avg}, a strong agreement between $c_{\alpha 2}$ based on H$_2$ and O$_2$ mass fractions can be observed. This suggests that the introduction of mixture fraction variation results in coupling between these two normalised species mass fractions. In contrast to $c_{\alpha 1}$, which exhibits super-unity values for the $\mathrm{O_2}$ mass fraction in the positively curved regions, $c_{\alpha 2}$ based on O$_2$ mass fraction is bounded between $0.0$ and $1.0$ for both positively and negatively curved regions. \par
\begin{figure}[h!] 
  \centering
  \includegraphics[width=0.55\columnwidth]{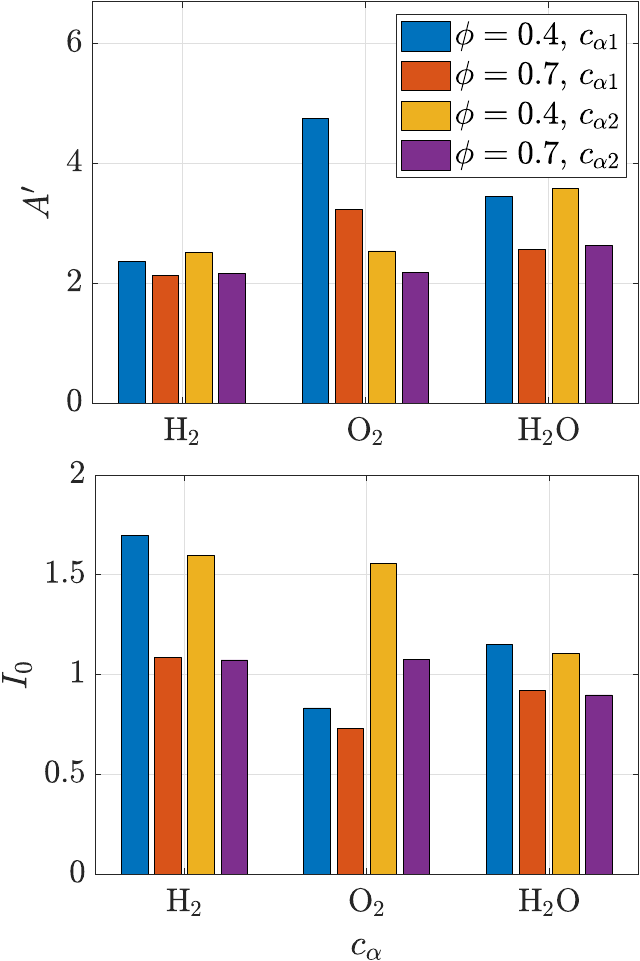}
  \vskip0.5em
  \caption{\footnotesize Variations of $A'$ and $I_0$ for different definitions of $c_{\alpha1}$ and $c_{\alpha2}$ for $\phi=0.4$ and $0.7$.}
  \label{bars}
\end{figure}

The differences in species distributions in response to the preferential diffusion effects induced by flame curvature are reflected in the differences in $c_{\alpha1}$ and $c_{\alpha2}$ distributions in Fig. \ref{c_avg} and Fig. \ref{c_avg_profiles}. This has implications on the evaluations of flame surface area and the stretch factor or efficiency factor, $I_0$, which signifies the ratio of reaction rate of the species in question per unit area normalised by the corresponding 1D unstretched laminar premixed flame value \cite{Bray1990}. For the present configuration, the stretch factor is computed using the relation: $S_f=S_L I_0 A'$, where $S_f$ is the flame propagation speed defined as per Eq. \ref{S_f}, and $A'$ represents the flame area enhancement, $A'=A_{\mathrm{wrinkled}}/A_{\mathrm{flat}}$. The wrinkled flame surface area $A_{\mathrm{wrinkled}}$ is evaluated as: $A_{\mathrm{wrinkled}} = \int_{0}^{Ly} \int_{0}^{Lx} |\nabla c_{i}| \ dx \ dy$ where $i=\alpha 1$ or $\alpha 2$ depending on $c_i$ definition. Figure \ref{bars} shows the values of $A'$ and $I_0$  calculated for both $c_{\alpha1}$ and $c_{\alpha2}$ at both equivalence ratios.

Moderate differences are observed in the values calculated for flame area enhancement and stretch factor for the different $c_{\alpha}$ definitions. It should be noted here that there are differences in the values for $A'$ depending on if $c_{\alpha 1}$ or $c_{\alpha 2}$ are used for the evaluation. The flame area enhancement, $A'$, based on the ${\mathrm{O_2}}$ mass fraction assumes the highest value when $c_{\alpha 1}$ is considered for both equivalence ratios, whereas the smallest $A'$ values are obtained for ${\mathrm{H_2}}$ mass fraction when $c_{\alpha 1}$ is used for both equivalence ratios. In the case when $A'$ is based on $c_{\alpha 2}$, the overall trends for $A'$ across equivalence ratios remain the same, with an increase in area can be seen as $\phi$ decreases, but a notable reduction in $A'$ is observed for the $A'$ values using ${\mathrm{O_2}}$ based definition of $c_{\alpha 2}$, compared with $c_{\alpha 1}$ values. For all definitions of $c_{\alpha1}$ and $c_{\alpha2}$, area enhancement is greater for the $\phi$ = 0.4 case than in the $\phi$ = 0.7 case, which is a consequence of enhanced flame wrinkling in the $\phi$ = 0.4 case as a result of strong thermo-diffusive instability. However, the stretch factor $I_0$ exhibits an opposite trend to those seen for $A'$ with the ${\mathrm{H_2}}$ mass fraction based $c_{\alpha 1}$ yielding the highest $I_0$ values and the lowest values are obtained for ${\mathrm{O_2}}$ based $c_{\alpha 1}$. In the case of $c_{\alpha 2}$, ${\mathrm{O_2}}$ based definition yields significantly larger values of $I_0$ for both equivalence ratios, compared with $c_{\alpha 1}$. This implies that the influence of mixture fraction variation is significant on $I_0$ and this should be accounted for in the modelling strategies. Moreover, $I_0$ values for the $\phi=0.4$ case are found to be greater than the corresponding values for $\phi=0.7$ (e.g., $I_0$ values range between 0.83 and 1.70 for the $\phi=0.4$ case and between 0.73 and 1.09 for the $\phi$ = 0.7 case.), indicating that the influence of flame stretch strengthens with decreasing equivalence ratio $\phi$ in the case of lean ${\mathrm{H_2}}$-air premixed flames.

\section{Summary and Conclusions\label{sec:conc}} \addvspace{10pt}
High-fidelity 2D simulations of lean ${\mathrm{H_2}}$-air premixed flames subjected to perturbations have been conducted for equivalence ratios of $\phi=0.4$ and $0.7$ to analyse the influence of preferential diffusion effects on the distributions of major species and their reaction rates. It has been found that the preferential diffusion effects give rise to significant deviations of mass fractions of major species and their reaction rates from the corresponding 1D unstretched laminar premixed flame solution and this tendency is particularly strong for ${\mathrm{H_2}}$ among all the major species. These deviations arise principally due to preferential diffusion effects induced by relative focussing/defocussing of species and heat at the positively and negatively curved regions with the nature of the deviations being opposite to each other depending on the sign of the curvature. This is reflected in the local increases (decreases) in $\phi$ in the positively (negatively) curved zones in the 2D $\phi=0.4$ case. These trends strengthen and enhancements of burning rate, flame surface area, and stretch factor are observed with a decrease in equivalence ratio as the flame becomes increasingly susceptible to preferential diffusion and thermo-diffusive instability. It has been found that the likelihood of super-adiabatic temperature increases with a decrease in  $\phi$, and super-adiabatic values of ${\mathrm{O_2}}$ and ${\mathrm{H_2O}}$ are obtained in the super-adiabatic temperature zones, especially for the $\phi=0.4$ case. This suggests that ${\mathrm{O_2}}$ and ${\mathrm{H_2O}}$ mass fraction based reaction progress variables are likely to be problematic for highly lean ${\mathrm{H_2}}$-air premixed flames because the reaction progress variable values for these definitions are not going to be bounded between $0.0$ and $1.0$. However, $\mathrm{O_2}$ mass fraction based reaction progress variable remains bound between $0.0$ and $1.0$ when the equivalence ratio variation is taken into account, but $\mathrm{H_2O}$ mass fraction based reaction progress variable shows increased non-monotonic behaviour within the flame front when the local equivalence ratio variation is accounted for. The normalised species distributions analysed here provide the platform based on which an optimal choice of reaction progress variable can be produced for lean $\mathrm{H_2}$-air flames using automated methodologies (e.g., principal component analysis, computational singular perturbation method, etc.) in the future.

\section*{Acknowledgements}\label{Acknowledgements}
The authors are grateful for the financial support from Tony-Trapp PhD studentship (provided by Dr Tony Trapp) and the Engineering and Physical Sciences Research Council (Grant: EP/W026686/1). The computational support for this work was provided by ARCHER2 Pioneer projects (e691 \& e817) and ROCKET HPC facility at Newcastle University.









\bibliographystyle{elsarticle-num}
\bibliography{PCI_LaTeX}

\end{document}